\documentclass[preprint,review,12pt]{elsarticle}

\usepackage{graphicx}
\usepackage{float} %required for the placement specifier H
\usepackage{amssymb}
\usepackage{lineno}
\usepackage{xcolor}  % package for colored text
\usepackage{siunitx} % SI unit package
\usepackage{svrsymbols} %Table 336: svrsymbols Physics Ideograms
\usepackage[utf8]{inputenc} % Required for inputting international characters
\usepackage[T1]{fontenc} % Output font encoding for international characters

\journal{Nuclear Inst.and Methods in Physics Research}

\begin{document}

\begin{frontmatter}

%% Title, authors and addresses

\title{Front-end ASIC for Spectroscopic Readout of Virtual Frisch-Grid CZT Bar Sensors}
%\title{Design and Characterization of a Multi-channel Application Specific Integrated Circuit for Spectroscopic Readout of 3D Position Sensitive Virtual Frisch-Grid Sensors}

%% use the tnoteref command within \title for footnotes;
%% use the tnotetext command for the associated footnote;
%% use the fnref command within \author or \address for footnotes;
%% use the fntext command for the associated footnote;
%% use the corref command within \author for corresponding author footnotes;
%% use the cortext command for the associated footnote;
%% use the ead command for the email address,
%% and the form \ead[url] for the home page:
%%
%% \title{Title\tnoteref{label1}}
%% \tnotetext[label1]{}
%% \author{Name\corref{cor1}\fnref{label2}}
%% \ead{email address}
%% \ead[url]{home page}
%% \fntext[label2]{}
%% \cortext[cor1]{}
%% \address{Address\fnref{label3}}
%% \fntext[label3]{}

%% use optional labels to link authors explicitly to addresses:
%% \author[label1,label2]{<author name>}
%% \address[label1]{<address>}
%% \address[label2]{<address>}

\author[mainaddress,collaboratoraddress]{Emerson Vernon\corref{correspondingauthor}}
\ead{evernon@bnl.gov, evernon@stonybrook.edu}

\author[collaboratoraddress]{Gianluigi De Geronimo\corref{correspondingauthor}}
\ead{degeronimo@ieee.org}
\author[mainaddress]{Aleksey Bolotnikov}
\author[collaboratoraddress]{Milutin Stanacevic}
%\author[mainaddress]{Abdul Rumaiz}
\author[mainaddress]{Jack Fried}
\author[mainaddress2]{Luis Ocampo Giraldo}
\author[mainaddress]{Graham Smith}
\author[mainaddress]{Kevin Wolniewicz}
\author[mainaddress]{Kim Ackley}
\author[mainaddress1]{Cynthia Salwen}
%\author[mainaddress]{John Triolo}
\author[mainaddress]{John Triolo\fnref{jt}}
\fntext[jt]{Deceased}
\author[mainaddress]{Don Pinelli}
\author[mainaddress]{Kenny Luong}

\cortext[correspondingauthor]{Corresponding author}

\address[mainaddress]{Instrumentation Division, Brookhaven National Laboratory, New York, United States}
\address[mainaddress1]{Environmental and Climate Sciences Department, Brookhaven National Laboratory, New York, United States}
\address[collaboratoraddress]{Department of Electrical \& Computer Engineering, Stony Brook University, New York, United States}
\address[mainaddress2]{Idaho National Laboratory, Idaho, United States}

% https://arxiv.org/
\begin{abstract}
%% Text of abstract

Compact multi-channel radiation detectors rely on low noise front-end application specific integrated circuits (ASICs) to achieve high spectral resolution. Here, a new ASIC developed to readout virtual Frisch-grid cadmium zinc telluride (VFG CZT) detectors for gamma ray spectroscopy is presented. Corresponding to each ionizing event in the detector, the ASIC measures the amplitude and timing at the anode, the cathode and four pad sense electrodes associated with each sensor in a detector array. The ASIC is comprised of 52 channels of which there are 4 cathode channels and 48 channels which can be configured as either anode channels with a baseline of 250 mV or pad sense channels to process induced signals with a baseline of 1.2 V.  With a static power dissipation of 3 mW, each channel performs low-noise charge amplification, high-order shaping, peak and timing detection along with analog storage and multiplexing. The overall channel linearity was better than $\pm$ 1 \si{\percent} with timing resolution down to 700 ps for charges greater than 8 fC in the 3 MeV range. With a 4 x 4 array of 6 x 6 x 20 \si{\cubic\milli\meter} virtual Frisch-grid bar sensors connected and biased, an electronic resolution of $\approx$ 270 \si{\electron} rms for charges up to 100 fC in the 3.2 MeV range was measured. Spectral measurements obtained with the 3D correction technique demonstrated resolutions of 1.8 \si{\percent} FWHM at 238 keV and 0.9 \si{\percent} FWHM at 662 keV.
\end{abstract}

\begin{keyword}
Application Specific Integrated Circuit (ASIC) \sep low noise front-end \sep mixed signal \sep Cadmium Zinc telluride (CdZnTe or CZT) \sep thallium bromide (TlBr) \sep mercuric iodide (HgI$_{2}$) \sep virtual Frisch-grid \sep 3D  \sep gamma ray \sep spectroscopy, radiation detection
%% keywords here, in the form: keyword \sep keyword
%% MSC codes here, in the form: \MSC code \sep code
%% or \MSC[2008] code \sep code (2000 is the default)
\end{keyword}

\end{frontmatter}

%% Start line numbering here if you want
%\linenumbers

%% main text
\section{Introduction}
\label{Introduction}

The demand for compact, high resolution and low power multi-channel radiation detection systems to explore the frontier of nuclear science brings together a group of stakeholders with application interest in security, medicine, industry, geology, space exploration and high energy physics (HEP). At the forefront of the technological innovation space are advances in material science and monolithic application specific integrated circuits (ASICs). We developed an ASIC prototype that was optimized for virtual Frisch-grid cadmium zinc telluride (VFG CZT) sensors. Given the interest in room temperature detectors, we instrumented additional circuitry on the chip to facilitate proof of concept designs with other materials such as Mercuric iodide (HgI$_{2}$) and thallium bromide (TlBr) sensors.

Gas, liquid, scintillator, and semiconductor sensors have been successfully demonstrated for x- and $\gamma$-ray spectroscopy \cite{Smith:1997GammaRayD} with niche applications based on practical trade-offs between cost, detection efficiency and achievable resolution. High purity germanium (HPGe) has the best energy resolution for sources with closely spaced energy peaks and low energy photons but these sensors must be cooled to temperatures around 130-77 K to yield resolutions down to $\approx$ 0.3$\%$ at 662 keV \cite{knoll:2000,Cooper:2015nim} which constrain their deployment as portable radiation detectors.

Wide bandbap semiconductors such as mercuric iodide (HgI$_{2}$), thallium bromide (TlBr), and cadmium zinc telluride (CZT) are three attractive room temperature sensor materials with low dark current and high photoelectric cross-section. Polarization and charge trapping degrade the spectral performance of HgI$_{2}$ \cite{Hermon:1996nim}, yet, detectors with single pixel resolution below 1$\%$ FWHM at 662 KeV were reported \cite{Meng:2006tns}. TlBr is attracting renewed interest for gamma ray spectroscopy \cite{Kim:2011nim,Churilov:2009tns} as techniques are being developed to address polarization in biased sensors \cite{Kozorezov:2010jap,Thrall:2012nssmic,HITOMI:2008nim,Datta:2017apl,Datta:2018tns}. Single pixel resolution on the order of 1$\%$ FWHM at 662 keV was reported at 20 \si{\degreeCelsius} where the detector showed stable performance \cite{Onodera:2004nim,Koehler:2015tns}. CZT has become the choice material for room temperature gamma ray detection and spectroscopy\cite{Eisen:1998jcg,Takahashi:2001tns}. Improvements in growth and characterization techniques are trending towards better spectral performance \cite{Sordo:2009sensors,BOLOTNIKOV:2012nim}. 

For high energy gamma ray spectroscopy, large volume premium CZT crystals are required to achieve good detection efficiency and resolution < 1 \si{\percent} FWHM at 662 keV. Since the yield on large volume premium crystals is impacted by crystal defects, smaller standard crystals are an attractive cost-effective alternative. To improve the resolution of CZT detectors, a number of techniques have been developed \cite{Zhang:2013sensors}. The fraction of the signal contribution due to hole trapping can be measured electronically then subtracted using the coplanar grid (CPG) technique \cite{He:2000nss,DeGeronimo:2006nss}. Alternatively, electrostatic shielding using the small pixel effect 3D position sensing method \cite{HE:1999nim,He:1997nim,Zhang:2004tns,Zhangb:2004tns} or the Frisch-ring approach \cite{Bunemann:1949cjr,Dmitrenko:1982,Parnham:2000jcg,Szeles:2006spie,Montemont:2001tns,mcgregor:2001pat} can make the crystal single polarity ($\electron$) sensing for improved performance. In a compact multi-channel system that utilizes these capabilities, custom ASICs are required for low noise and low power signal processing. The Virtual Frisch-Grid (VFG) approach incorporated an optimization of the Frisch-ring technique \cite{POLACK:2010nim,Bolotnikov:2014pat,Bolotnikov:2012tns,Bolotnikov:2007spie}. This economical approach to CZT spectroscopy allows for the development of commensurable large volume detectors by tiling unit bar sensors (up to 40 mm thick) made from standard grade material into large arrays \cite{Giraldo:2017tns}.

\begin{figure}[!t]
	\centering\includegraphics[width=1.0\linewidth]{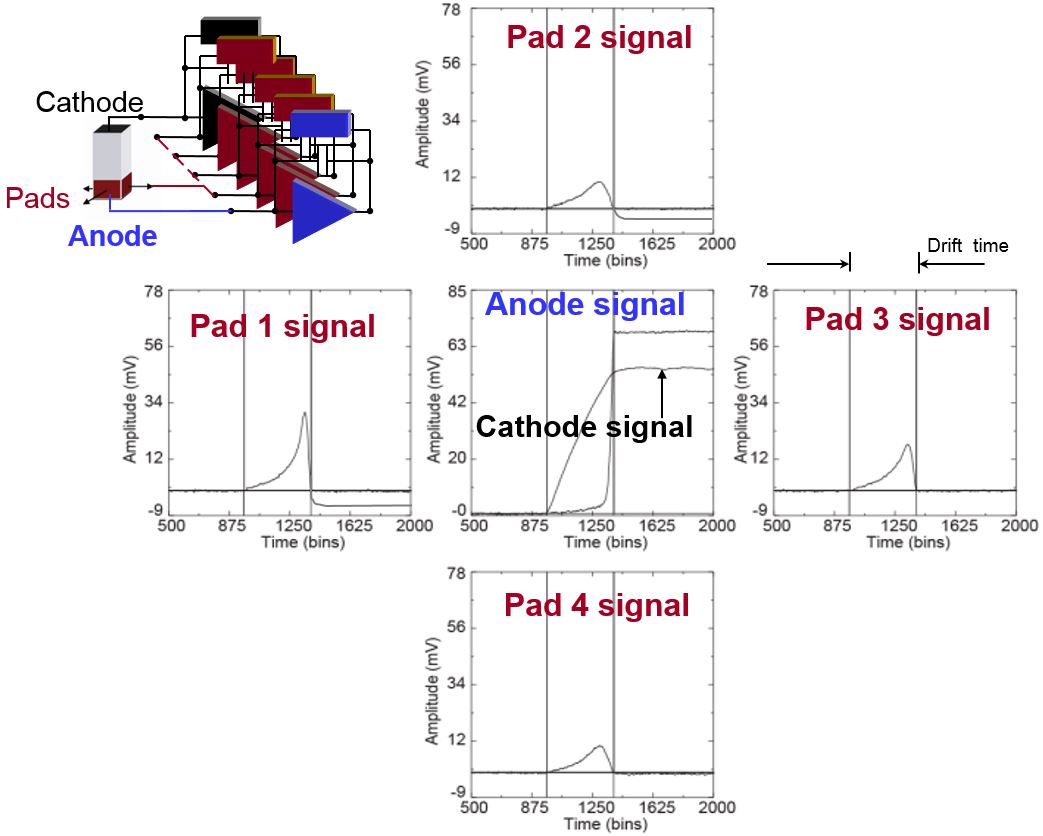}
	\caption{Simulated preamplifier output corresponding to charge collected on six electrodes (anode, cathode and four pads) of a Virtual Frisch-grid CZT bar sensor.}
	\label{fig:vfgIdlPampSim1}
\end{figure}

To facilitate practical applications of the VFG sensors, an advanced virtual Frisch-grid (AVG) ASIC was developed to independently readout the six electrodes (anode, cathode and four sense pads) associated with each VFG CZT bar sensor as shown in Figure \ref{fig:vfgIdlPampSim1}. Each channel in the ASIC performs charge preamplification, high order shaping with configurable baseline (including baseline stabilization), amplitude measurement with corresponding event time extraction, analog data storage and multiplexing. In section \ref{AVG ASIC Architecture}, the architecture of the ASIC is presented. This is followed by the experimental setup and results in sections \ref{Experimental Setup} and \ref{Experimental Results}. We close the paper with our conclusions and future work in section \ref{Conclusions and Future Work }.  

\section{AVG ASIC Architecture}
\label{AVG ASIC Architecture}

\subsection{Chip Level Architecture}

\begin{figure}[!t]
	\centering\includegraphics[width=1.0\linewidth]{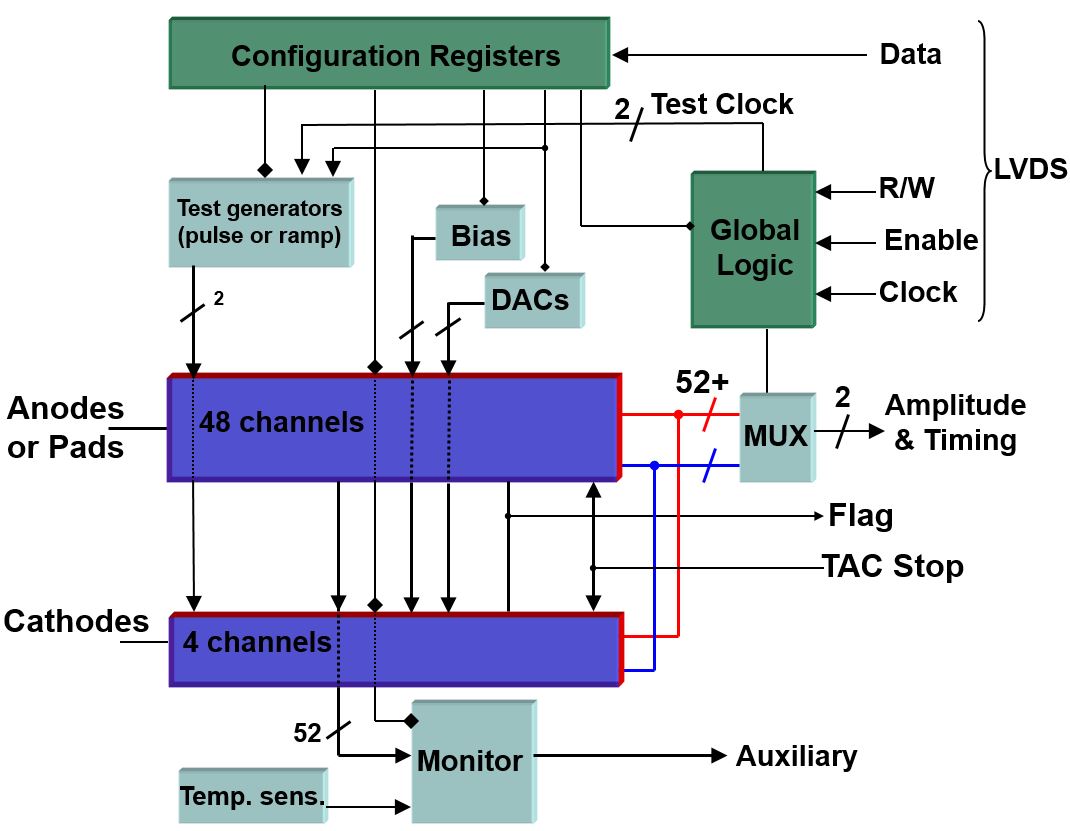}
	\caption{Block diagram of the second AVG ASIC prototype. The chip implements a global control logic, configuration registers, 52 programmable input channels, two test generators, a bias, DACs, global logic, analog multiplexers and a monitor for in-situ tracking of critical circuit blocks.}
	\label{fig:avg_block_diagram}
\end{figure}

The first AVG ASIC prototype designed to readout a 3 x 3 array of VFG CZT sensors has 36 anode channels to readout negative charge and 9 cathode channels to readout positive charge with a fixed 250 mV baseline. This architecture is upgraded to the second AVG ASIC prototype illustrated in Figure \ref{fig:avg_block_diagram} with changes to the channel configuration. To accommodate a 4 x 4 array of VFG bar sensors, the ASIC implements a total of 52 channels. The modular design incorporates 4 cathode channels (one for each 2 x 2 unit array of sensors) while each of the remaining 48 channels has the option to be independently configured as either an anode channel with a baseline of $\approx$ 250 mV or a pad sensing channel with a baseline of $\approx$ 1.2 V. 

A global logic selects between three modes of operation referred to as write-in, signal acquisition and readout. During write-in, a pseudo serial peripheral interface (pSPI) is used to shift data into the configuration registers from an external data acquisition system (DAQ). The configuration bits set the initial condition for the bias, adjust the digital-to-analog converters (DACs), and program the 52 channels and the test generators. The global bias block establishes the bias voltages and currents while the DACs set the threshold voltages for pulse amplitude discrimination and the output voltage of two test generators. In signal acquisition mode, the embedded generators are used to inject charge into the front-end.  

The analog signal from each critical ASIC block can be multiplexed and monitored on a dedicated auxiliary port while running diagnostics or evaluating the chip in-situ. All anode channels are optimized for 3.3 pF of input capacitance while the cathode channels are optimized for 6.3 pF of input capacitance \cite{DeGeronimo:2005tns}. The ASIC has four programmable gains from 120 mV/fC to 20 mV/fC covering the energy range from 530 KeV to 3.2 MeV in CZT. Eight programmable shaping times from 250 \si{\nano\second} to 12 $\mu$s are implemented to add flexibility to the design. The digital interface is low voltage differential signaling (LVDS) with on-chip 100 $\Omega$ termination resistors. Whenever a sensor event goes above threshold, the ASIC releases a flag. Subsequent readout by the external DAQ is realized with a token passing scheme. That is, the token controls the switching of the analog multiplexers, so only the channel in which the token resides can put analog data on the respective amplitude and timing ports for analog-to-digital conversion. In addition, a temperature sensor is integrated and its analog data is multiplexed onto the tail of the data stream after the last channel is read.

\subsection{Anode and Pad Sensing Channels}
\label{Anode and Pad Sensing Channels}

\begin{figure}[!t]
	\centering\includegraphics[width=1.0\linewidth]{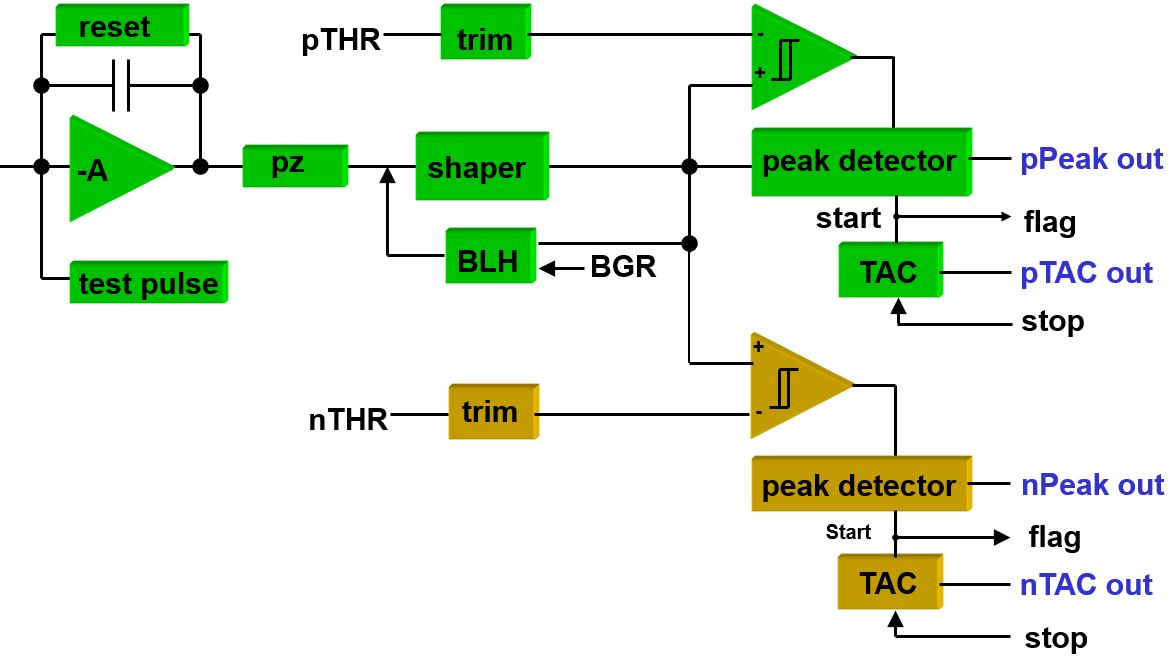}
	\caption[AVG anode and pad channel block diagram]{Block diagram of a channel that can be configured as an anode signal processing channel with a baseline of $\approx$250 mV or a pad sensing channel with a baseline of $\approx$1.2 V}
	\label{fig:AVGanodeBlockDiagram}
\end{figure}

A representation of the configurable anode or pad channel is presented in Figure \ref{fig:AVGanodeBlockDiagram}. The signal chain in each of the 48 channels is comprised of a charge preamplifier with pole-zero cancellation, a fifth order shaper with bandgap referenced baseline holder \cite{Geronimo:2000tnsbln} and a pair of extractor (discriminator, peak detector and time-to-amplitude converter)\cite{DeGeronimo:2002nimpt1,DeGeronimo:2002nimpt2,Dragone:2005nss}.  In anode mode, the total channel gain is configured as 20 mV/fC, 40 mV/fC, 60 mV/fC or 120 mV/fC while in pad sensing mode, the selectable channel gains are 13.3  mV/fC, 26.6 mV/fC, 40 mV/fC and 80 mV/fC. 

The input of the charge preamplifier implements a PMOS transistor that is optimized for 3.3 pF total capacitance at a drain bias current of 120 \si{\micro\ampere} \cite{DeGeronimo:2005tns}. At the optimization point, a gate geometry (W/L) of 310 \si{\micro\meter}/0.36 \si{\micro\meter} with a corresponding gate capacitance (C$_{g}$) of approximately 816 fF and a transconductance (g$_{m}$) of about 2.2 mS is obtained. The transistor contributes 79 rms electrons to the ENC at 1 \si{\micro\second} shaping time and dissipates $\approx$ 300 \si{\micro\watt}. In addition, a current mirror continuous adaptive reset connected in the feedback loop establishes the amplifier bias point and provides continuous reset of the circuit after each measurement \cite{Geronimo:1999nss,Geronimo:2000tns,DeGeronimo:2004tns}. The pole introduced by the feedback loop is cancelled with a scaled replica of the mirror which created a zero at the output of the preamplifier. With this arrangement, the charge amplifier provides selectable charge gains of 48 or 96 \cite{DeGeronimo:2007nss}. 

\begin{figure}[!t]
	%\centering\includegraphics[scale=0.4]{figures/shaperResp2j}
	\centering\includegraphics[width=1.0\linewidth]{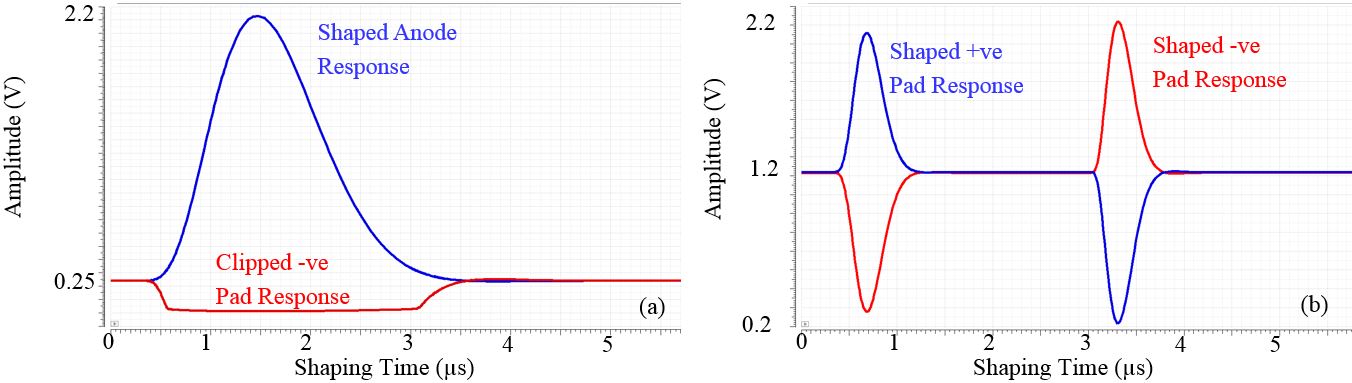}	
	\caption{Simulation of a channel in: (a) anode configuration with a baseline of 250 mV for unipolar pulse shaping. If connected to a sensor pad electrode, the negative pulse response to induced pad transients are clipped (b) pad configuration with a baseline of 1.2 V added in the second prototype. Full amplitude response to induced transient from the pad electrode.}
	\label{fig:shaperResp}
\end{figure}

Signal shaping is realized with a 5$^{th}$ order complex conjugate semi-Gaussian shaper \cite{DeGeronimo:2011tns} with adjustable shaping times of 250 ns, 500 ns, 1 \si{\micro\second} and 2 \si{\micro\second}. Cognizant of the developments in wide bandgap sensors such as TlBr, additional circuitry is implemented for selectable shaping times of 3 \si{\micro\second}, 6 \si{\micro\second} and 12 \si{\micro\second} to facilitate proof of concept designs with sensors characterized by longer drift times. The simulated shaper output in anode configuration and pad configuration is shown in Figures \ref{fig:shaperResp}a and \ref{fig:shaperResp}b respectively. 

\begin{figure}[!t]
	%\centering\includegraphics[scale=0.4]{figures/shaper2j}
	\centering\includegraphics[width=1.0\linewidth]{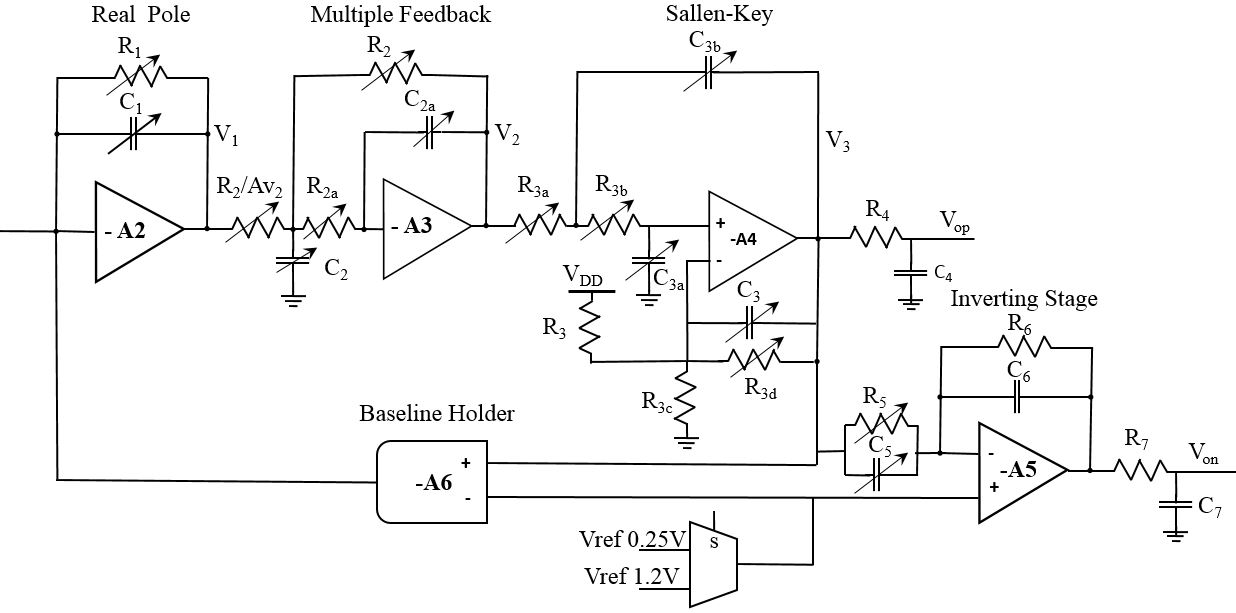}
	\caption{A 5$^{th}$ order shaper with one real pole, two pairs of complex conjugate poles, a baseline holder and inverting stage with programmable bandgap references of 0.25 V and 1.2 V.} 
	\label{fig:shaper}
\end{figure}

The first ASIC prototype implements 45 channels with fixed baselines of $\approx$ 250 \si{\milli\volt} for shaper responses up to 2 V above baseline as shown by the blue trace in Figure \ref{fig:shaperResp}a. When connected to a 3 x 3 array of VFG sensors, this ASIC can readout charge from 9 cathodes, 9 anodes and 27 pads. For the channels that are connected to the sensor pad electrodes, the shaped negative pulse response to an induced transient is clipped as shown by the red trace in Figure \ref{fig:shaperResp}a since there is not enough headroom for the negative voltage swing. 

In the second prototype, solving this problem by designing a dedicated channel to process the transient pad signals would almost double the power and the silicon area. This issue is addressed in Figure \ref{fig:shaper} by adding an analog inverting stage after the shaper and shifting the baseline to 1.2 V when a channel is configured to process pad induced signals with the response shown in Figure \ref{fig:shaperResp}b. Forty eight of these configurable channels are implemented in the second prototype.

Pulse height discrimination and threshold equalization is realized by two low hysteresis discriminators with 5 bit trims per channel that compensates for baseline dispersions. A pair of peak detector and time-to-amplitude converters (TACs) extracts and store the bi-parametric amplitude and analog timing data corresponding to an event.

\subsection{Cathode Channel}

\begin{figure}[!t]
	\centering\includegraphics[width=1.0\linewidth]{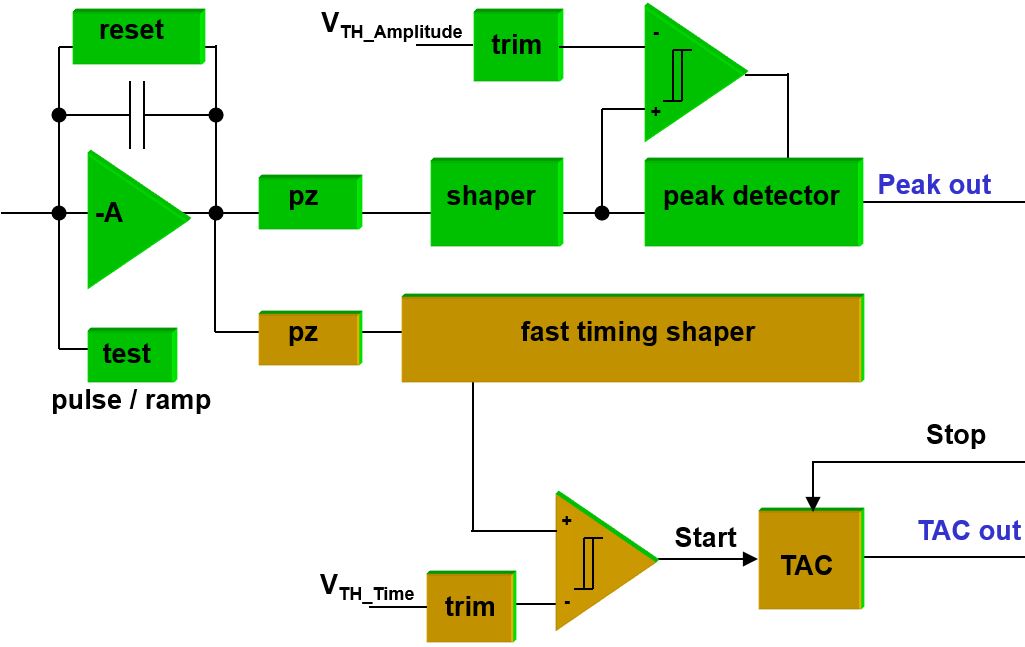}
	\caption{AVG cathode channel block diagram. The signal path splits at the output of the preamplifier. The event amplitude (Peak out) is processed by the green signal path while the corresponding timing (TAC out) is processed by a fast 3$^{rd}$ order shaper and a time-to-amplitude converter (TAC).}
	\label{fig:AVGcathodeBlockDiagram}
\end{figure}

The chip is designed with four cathode channels that service a modular 4 x 4 array of sensors. Each of the four cathode channels independently processes positive charge from a cathode electrode that is common to a unit 2 x 2 sensor array. A block diagram of the cathode channel is illustrated in Figure \ref{fig:AVGcathodeBlockDiagram}. The channel implements a low noise charge preamplifier optimized for 6.3 pF input capacitance, a pulse or ramp test circuit and a feedback network that reset the circuit after each measurement. 

The output of the preamplifier is split into two parallel signal processing chains. One signal path measures the event amplitude with selectable channel gain of 20 mV/fC or 60 mV/fC while the other signal path measures the event timing with selectable gain of 27 mV/fC and 162 mV/fC. The amplitude-processing branch was similar to the anode channel presented in section \ref{Anode and Pad Sensing Channels}.

The event timing is realized as a fast 3$^{rd}$ order unipolar shaper which comprises one real pole and a pair of complex conjugate poles with adjustable peaking times of 100 ns, 200 ns, 400 ns, and 800 ns. A band-gap referenced baseline holder prevents baseline drift while a threshold discriminator with 4-bit trimming triggers a programmable time-to-amplitude converter (TAC) for above threshold events. The trigger from the discriminator initiates a voltage ramp that is sampled by a reference signal from the external DAQ. The sampled voltage is stored on analog memory inside the TAC for readout.

\subsection{ASIC Prototype}

\begin{figure}[!t]
	\centering\includegraphics[width=0.95\linewidth]{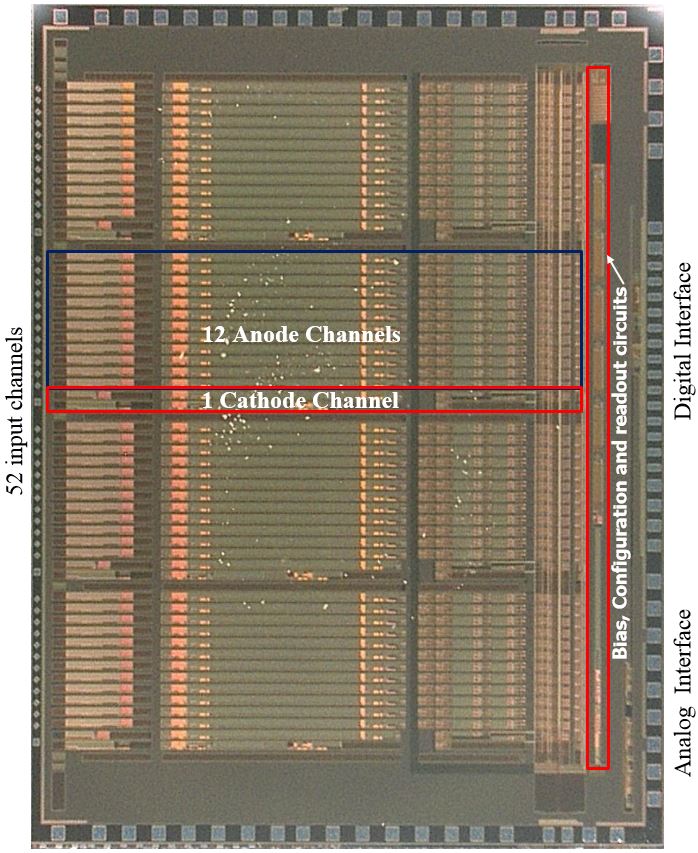}
	\caption{A photograph of the second AVG ASIC prototype. There are 52 input channels arranged in four groups of 12 anodes and a cathode.  At the back-end are bias, configuration, readout control and interface circuity. The layout size is 7.6 mm x 10 mm.}
	%\caption{Fabricated second AVG ASIC prototype with 48 configurable channels for anode signals or pad signals and 4 cathode channels to read out a 4 x 4 array of VFG CZT sensors. The chip implements four groups of 12 anodes and a cathode. Each channel is arranged with a preamplifier, a shaper with baseline holder, a pair of peak and time detectors and local logic with registers. The common circuits such as buffers, bias, DACs and global control logic with registers are at the back-end of the chip.}
	\label{fig:avg2Micrograph}
\end{figure}

A micrograph of the second ASIC prototype is shown in Figure \ref{fig:avg2Micrograph}. On the left side of the die are 52 input channels that interface with the sensor electrodes. The channels are arranged in four groups of twelve anode/pad channels plus a cathode channel. At the channel level, the functional blocks are linearly arranged with a preamplifier at the input followed by a shaper then two peak detectors, two time detectors and local control logic. The common circuits such as buffers, bias, DACs, configuration registers and global logic are at the back-end of the layout. Power and ground are supplied from the top and bottom of the chip while the back-end is reserved for bidirectional signal interfacing with the external DAQ. The chip was fabricated in a commercial 250 nm process with a silicon footprint of 7.6 mm x 10 mm and dissipates 177 mW of static power.

\section{Experimental Setup}
\label{Experimental Setup}
\begin{figure*}[!t]
	\centering\includegraphics[width=1\linewidth]{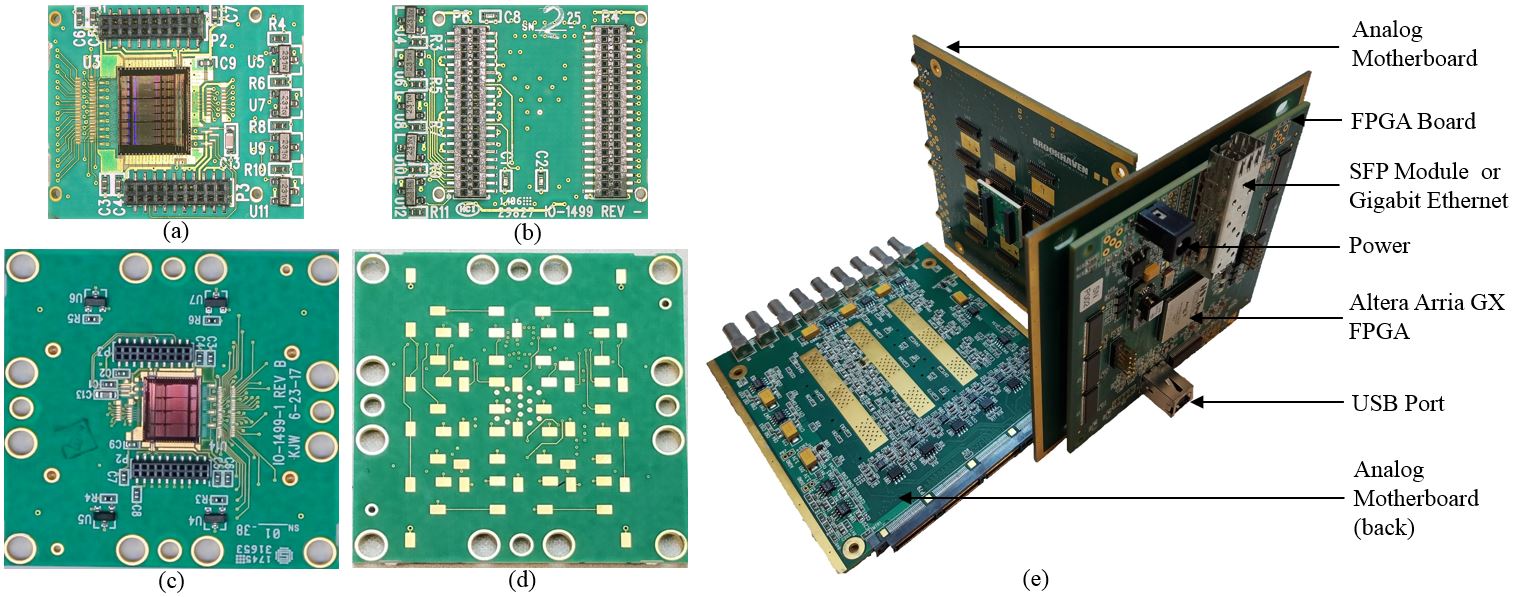}
	\caption{A 45 channel first ASIC prototype wirebonded to (a) the front of a 28 mm x 22 mm carrier board with (b) 2 connectors on the back to accommodate the sensor array. The second 52 channel ASIC prototype (c) mounted on the front of a 45 mm x 45 mm interposer while (d) the pads on the back interface with the detector module through a Z-ray connector. (e) An assembled external DAQ with the first ASIC carrier board mounted on an analog motherboard.}
	\label{fig:avg1Interposer}
\end{figure*}

The first prototype of the ASIC was wirebonded to a custom 28 mm x 22 mm carrier board shown in Figure \ref{fig:avg1Interposer}(a) with on board decoupling capacitors and discrete electrostatic discharge protection devices for the cathode channels. On the front of the carrier board were two electrically isolated connectors for digital (top) and analog (bottom) power and signal interface respectively. On the back (Figure \ref{fig:avg1Interposer}(b)) of the board were two connectors that accommodated a detector board which housed a 3 x 3 array of VFG CZT sensors.  Similarly, the 52 channel revised prototype was wirebonded to the front of a 45 mm x 45 mm interposer (Figure \ref{fig:avg1Interposer}(c)). The detector was connected to the back of the board (Figure \ref{fig:avg1Interposer}(d)) through a Z-ray connector \cite{Bolotnikov:2018nim}.  

A picture of the external DAQ is shown in Figure \ref{fig:avg1Interposer}(e) with one of two analog motherboards plugged into the assembly. A first prototype carrier board was mounted on the front of an analog motherboard which provides low noise (< 20 \si{\micro\volt} rms) power and analog-to-digital conversion as depicted by the discrete components on the back of the second motherboard. In the fully assembled system, the two motherboards were read out by an Altera Arria GX FPGA board that supplied system power along with the option for optical fiber, Gigabit Ethernet or USB communication. The system required only two connections to operate. That is, one connection for main power and the other for communication with a computer. During measurements, the system was placed in a slotted aluminum enclosure with thinned walls \cite{Ocampo:2018osti} that formed a Faraday cage around the ASICs and sensors. 

\section{Experimental Results}
\label{Experimental Results}
%Reference to Section \ref{Experimental Setup}. 

\begin{figure}[!t]
	\centering\includegraphics[width=1.0\linewidth]{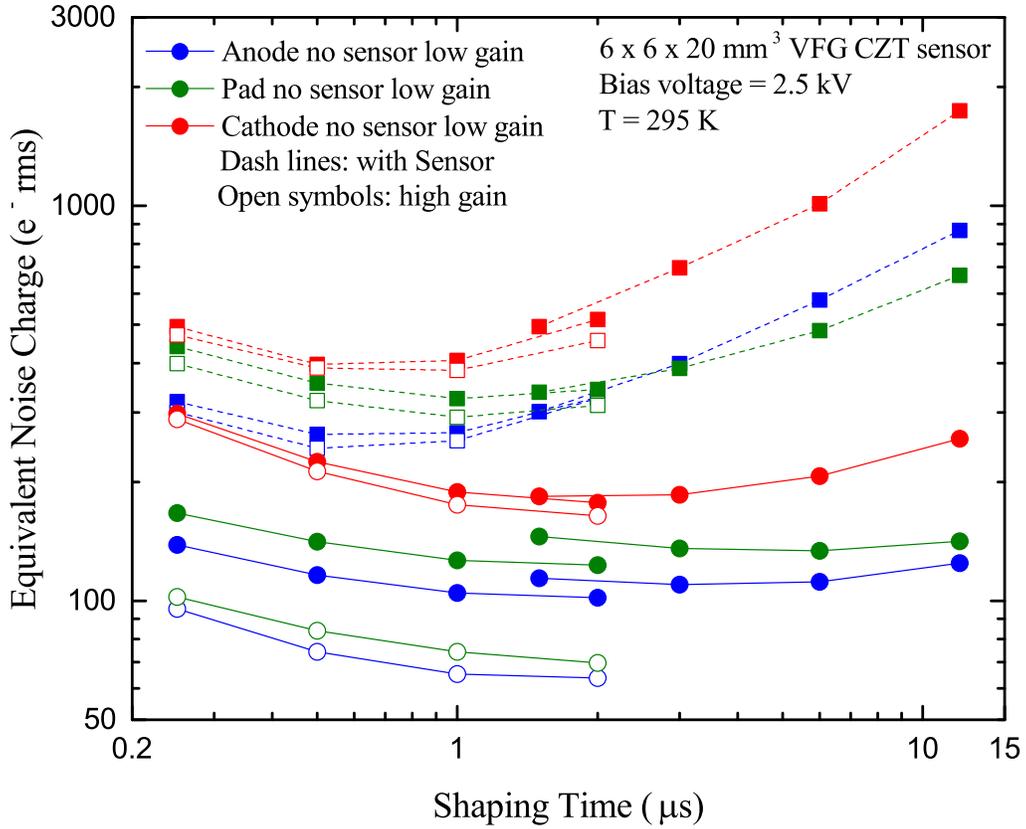}
	\caption{Measured equivalent noise charge at low gain (solid symbols) and high gain (open symbols) for an anode channel (blue traces), a pad sense channel (green traces) and a cathode channels (red traces) as a function of shaping time.The solid lines are measurements without a sensor and the dashed lines are measurements with a 6 x 6 x 20 \si{\cubic\milli\meter} VFG CZT sensor.}
	\label{fig:avg2Enc}
\end{figure}

Figure \ref{fig:avg2Enc} shows the measured equivalent noise charge (ENC) as a function of shaping time for an anode channel (blue), a pad channel (green) and a cathode channel (red) with and without the sensor connected. The noise voltage at the buffered output of the shaper was measured with a Rhode and Schwarz true RMS meter then referred to the input of the preamplifier. Without the sensor connected, at high gain covering the 530 keV energy window in CZT, the anode channel gave an ENC of $\approx$ 60 \si{\electron} RMS at 1 \si{\micro\second} shaping time for an effective dynamic range of $\approx$760. Further, at low gain (3.2 MeV maximum photon energy) an ENC of 100 \si{\electron} corresponding to a dynamic range of $\approx$ 2900 was attained. As designed, the difference is attributed to the noise contribution from the shaper. With a 6 mm x 6 mm x 20 mm VFG CZT senor connected and biased between 2-3 kV, the dynamic range achieved was $\approx$ 900 and $\approx$ 200 for low gain and high gain respectively. The measured ENC without the senor at each shaping time was less than that measured with the sensor.  This can be explained as follows. Without the sensor connect, the dominant part of the interconnect capacitance (on the order of a few hundred femto-Farads) was already incorporated in the carrier board. With the sensor connected, each electrode capacitance (on the order of pico-Farads) substantially increases the contribution
of the series noise (white and low-frequency). In addition, the sensor leakage current (bias voltage dependent) contributed quadratically to the total ENC. 

The cathode which was optimized for a larger capacitance and smaller gains (20 mv/fC and 60 mV/fC) had an electronic noise contribution that was less than 300 \si{\electron} rms at all shaping times without the sensor. With the sensor connected and biased (dotted traces), the ENC for the cathode was below 450 \si{\electron} rms at 1 \si{\micro\second} shaping time. For VFG CZT applications, the maximum shaping time did not exceed 2 \si{\micro\second}. At longer shaping times, the parallel noise dominated due to increased integration time of the current from the adaptive reset and the sensor leakage current. Though less than optimal for the ENC, we implemented the longer shaping times for prototyping and proof of concept designs with sensor material such as TlBr and HgI$_{2}$ that have long drift times. 

\begin{figure}[!t]
	\centering\includegraphics[width=1.0\linewidth]{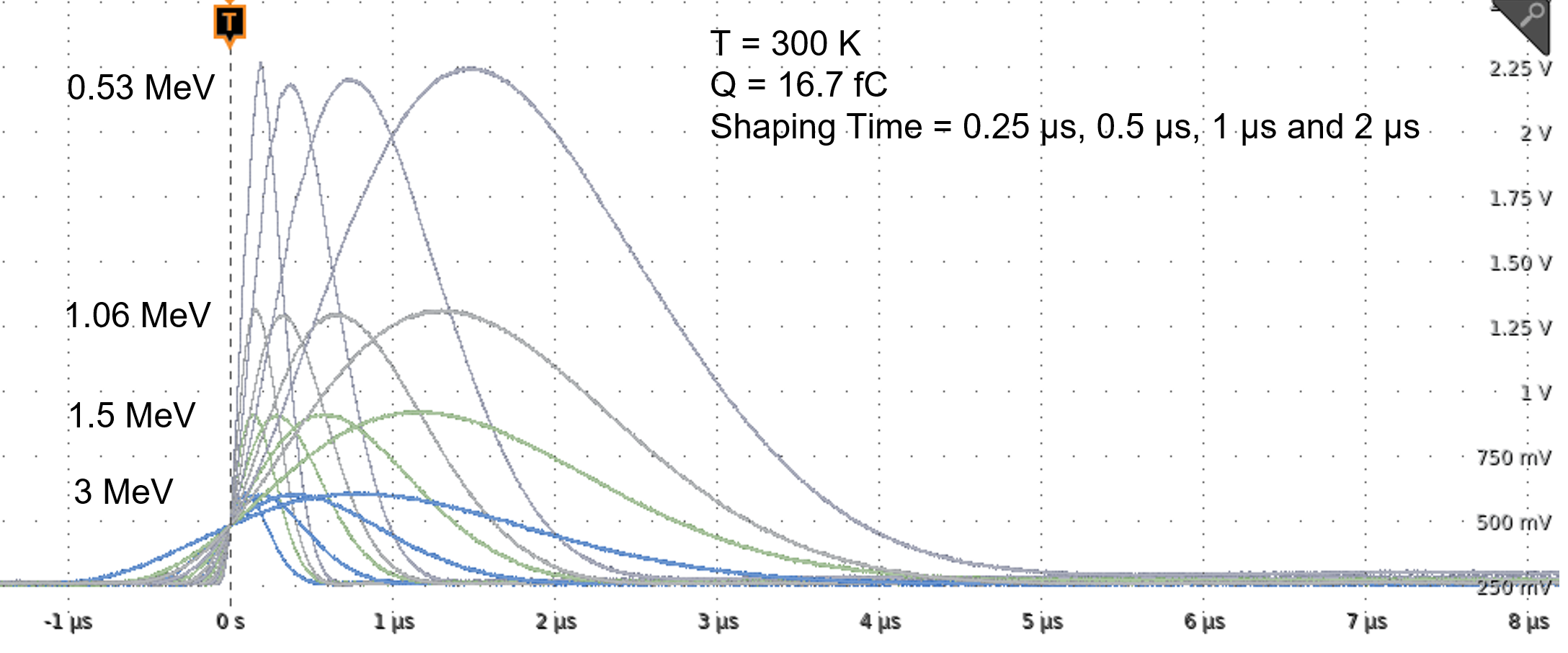}
	\caption{Measured anode response to 16.7 fC of charge at four gains and four shaping times}
	\label{fig:anodeMeasuredResponse}
\end{figure}

The anode channel response to 16.7 fC of injected charge from the on-chip test generator at four shaping times (0.25 \si{\micro\second}, 0.5 \si{\micro\second}, 1 \si{\micro\second} and 2 \si{\micro\second}) and four gains (covering 0.53 MeV, 1.06 MeV, 1.6 MeV and 3.2 MeV in CZT) is shown in Figure \ref{fig:anodeMeasuredResponse}. The anode had a baseline of 250 mV with a maximum voltage swing of $\approx$ 2 above baseline. 

\begin{figure}[!t]
	\centering\includegraphics[width=1.0\linewidth]{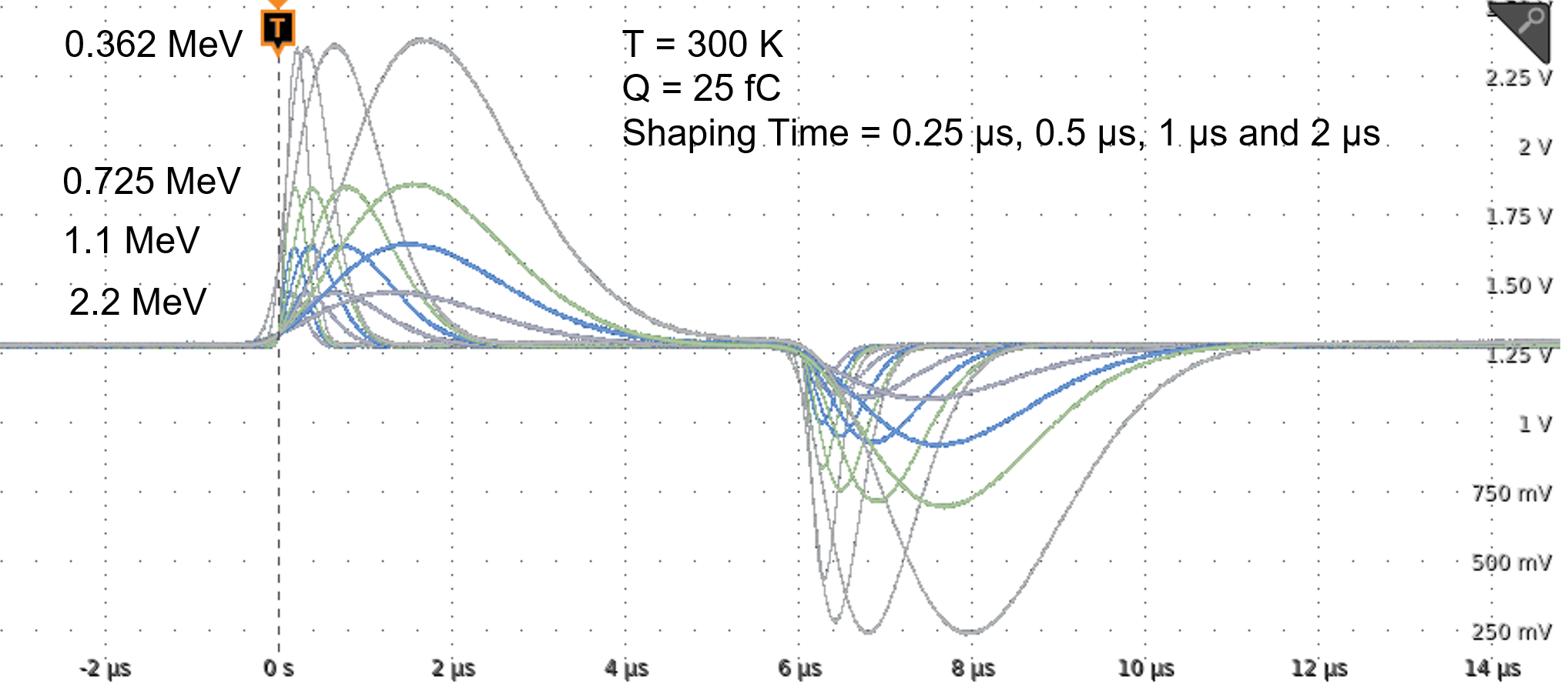}
	\caption{Measured sensing pad response to 25 fC of charge at four gains and four shaping times}
	\label{fig:padMeasuredResponse}
\end{figure}

Similarly, a plot of the same channel's response to 25 fC of injected charge when configured as a pad sensing channel to process induced charge is shown in Figure \ref{fig:padMeasuredResponse}. In this configuration, the baseline was held at 1.2 V and the measured shaper response to induced transients were bipolar pulses with amplitude swings up to $\pm$ 1 V. This was an improvement to \cite{DeGeronimo:2007nss,vernon:2010tns} that was designed to readout pixelated sensors with 250 mV baseline and maximum pulse amplitude responses of +2/-0.05 V.  During data acquisition, the ASIC measured the peak amplitude along with the corresponding timing data from the positive and negative peaks. This data was combined with the independent bi-parametric measurements from the other channels to correct the photopeak for each event.

\begin{figure}[!t]
	\centering\includegraphics[width=0.9\linewidth]{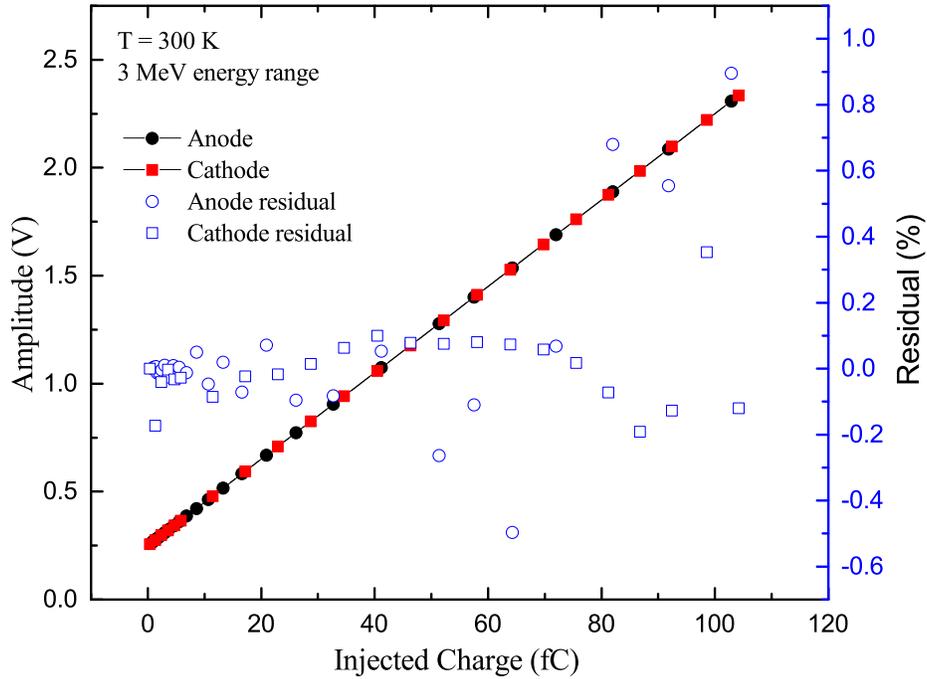}
	\caption{Measured anode and cathode linearity better than $\pm$ 1 \si{\percent} in the 3.2 MeV energy range for up to 100 fC of injected charge}
	\label{fig:avg2linearity}
\end{figure}

A plot of the best fit lines and the residuals for an anode and a cathode channel is shown in Figure \ref{fig:avg2linearity}. The anode had a linearity on the order of +0.9/-0.4 \si{\percent} while the cathodes were measured in the range of +0.4/-0.2 \si{\percent}. The cathode measurements were obtained by varying the amount of charge (up to 100 fC) injected into the front-end with the on-chip test generator and recording the corresponding amplitude. Regarding the anode on-chip test generator, a scaling factor in this circuit limited the maximum injected charge to 50 fC in the 3.2 MeV energy range. It should be noted that while this was a minor inconvenience for the evaluation of the chip, there was absolutely no impact on the designed charge measuring capabilities with a sensor connected. Therefore, the attenuated output of a LeCroy 9210 pulse generator was capacitively coupled to the anode front-end for the injection of charge (up to 100 fC).

The timing response of the anode channel was measured with externally injected charge while that of the cathode was evaluated with charge injected from the integrated test generator. For each increment of input charge (from 0.08 fC to 100 fC), representative samples of the time-to-amplitude converter (TAC) output were taken by the external DAQ. The statistical mean of the TAC output was used to evaluate the timewalk while the standard deviations were used to determine the time resolution. 

\begin{figure}[!t]
	\centering\includegraphics[width=0.66\linewidth]{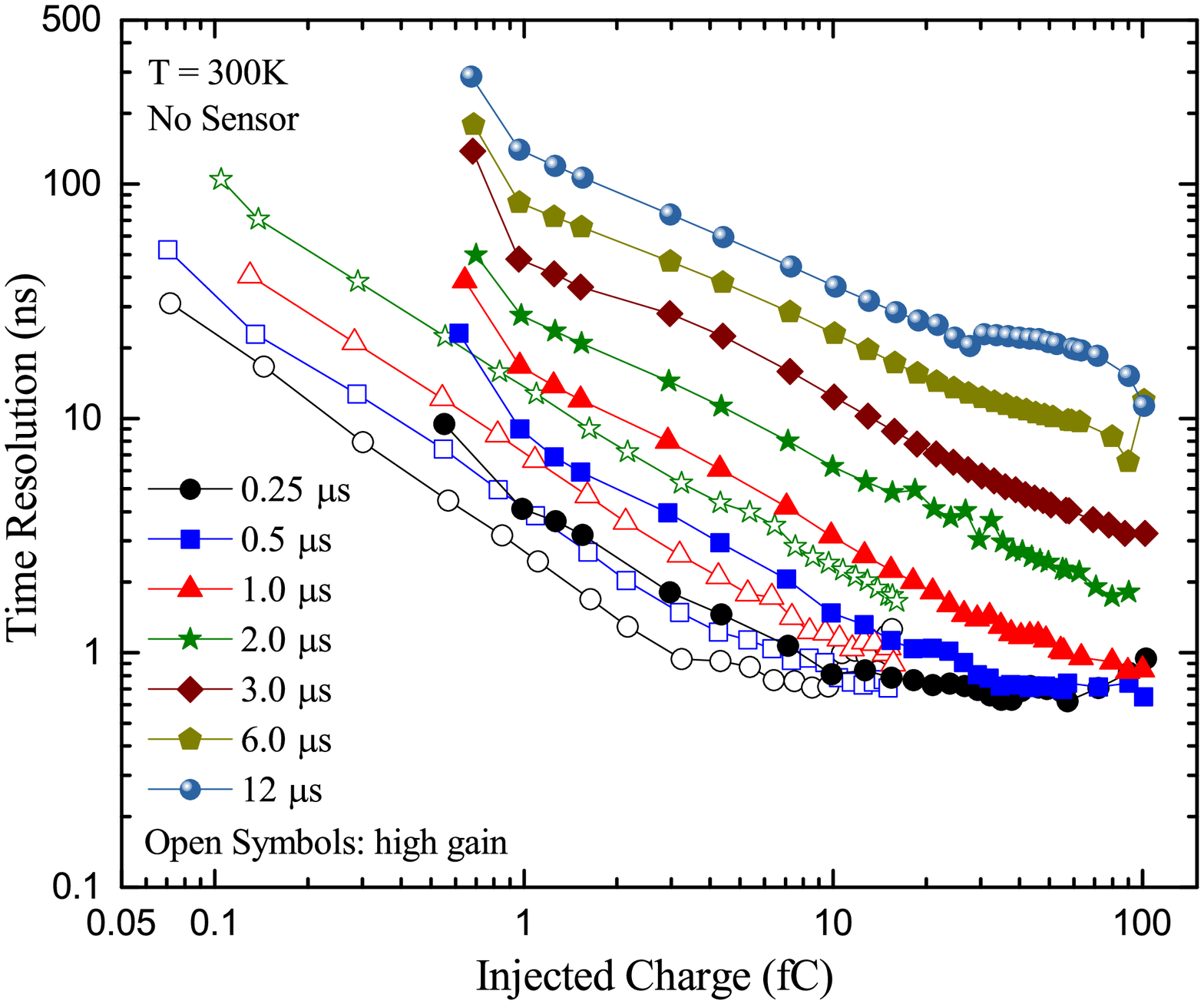}
	\caption{Measured anode time resolution at 20 mV/fC (solid symbols) and 120 mV/fC (open symbols) for injected charge up to 100 fC.}
	\label{fig:avg2AnodeTimeRes}
	
	\centering\includegraphics[width=0.66\linewidth]{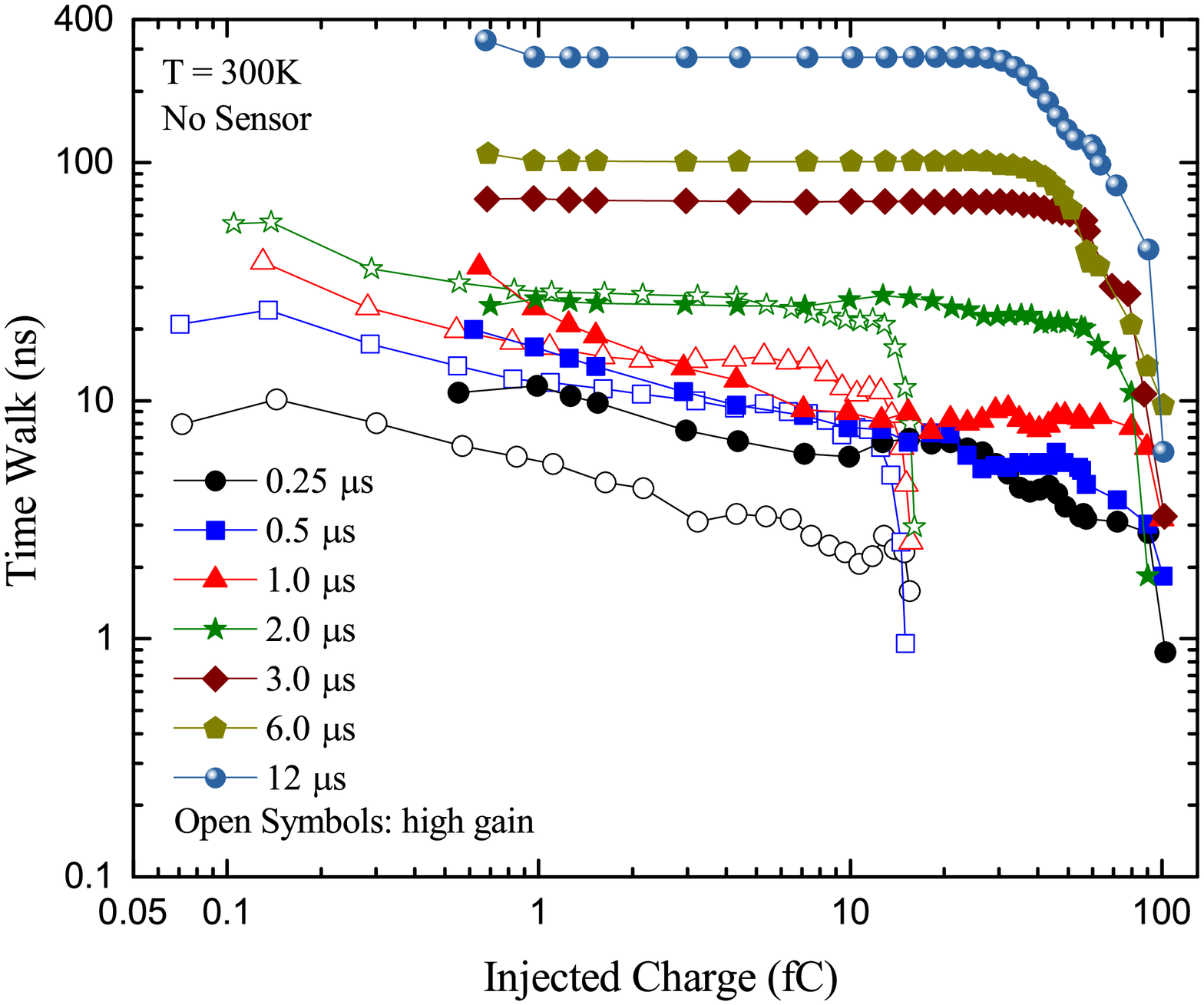}
	\caption{Measured anode time walk at 20 mV/fC (solid symbols) and 120 mV/fC (open symbols) for injected charge up to 100 fC.}
	\label{fig:avg2AnodeTimeWalk}
\end{figure}

Figure \ref{fig:avg2AnodeTimeRes} shows the anode time resolution. For a TAC duration of 1 \si{\micro\second}, a minimum time resolution of 700 ps rms was measured at high gain (120 mV/fC) and 0.25 \si{\micro\second} shaping when charge greater than 8 fC was processed. The ASIC demonstrated the capability to resolve charge down to 0.09 fC with a resolution of 31 ns. For a TAC duration of 2 \si{\micro\second}, low gain (20 mV/fC) and 12 \si{\micro\second} shaping time shaping time, input charges $\geq$ 0.7 fC were resolved at 290 ns or better. The corresponding timewalk adjusted to the minimum was captured in Figure \ref{fig:avg2AnodeTimeWalk} with a lower limit of 1.6 ns at high gain and 0.25 \si{\micro\second} shaping time to an upper limit of 330 ns at low gain and 12 \si{\micro\second} shaping time.  

\begin{figure}[!t]
	\centering\includegraphics[width=0.65\linewidth]{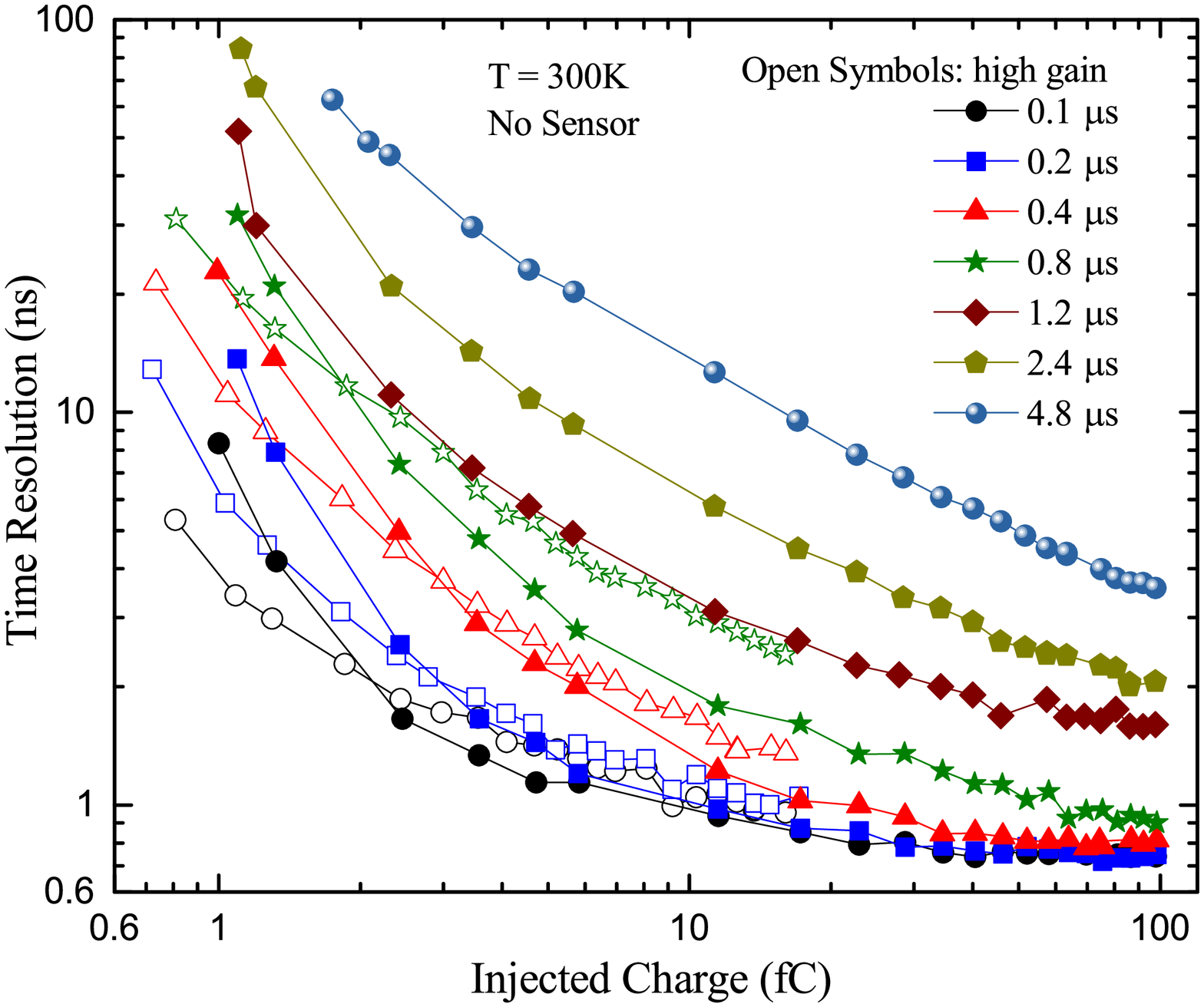}
	\caption{Measured cathode time resolution at 20 mV/fC (solid symbols) and 60 mV/fC (open symbols) for injected charge up to 100 fC.}
	\label{fig:avg2CathodeTimeRes}
	
	\centering\includegraphics[width=0.65\linewidth]{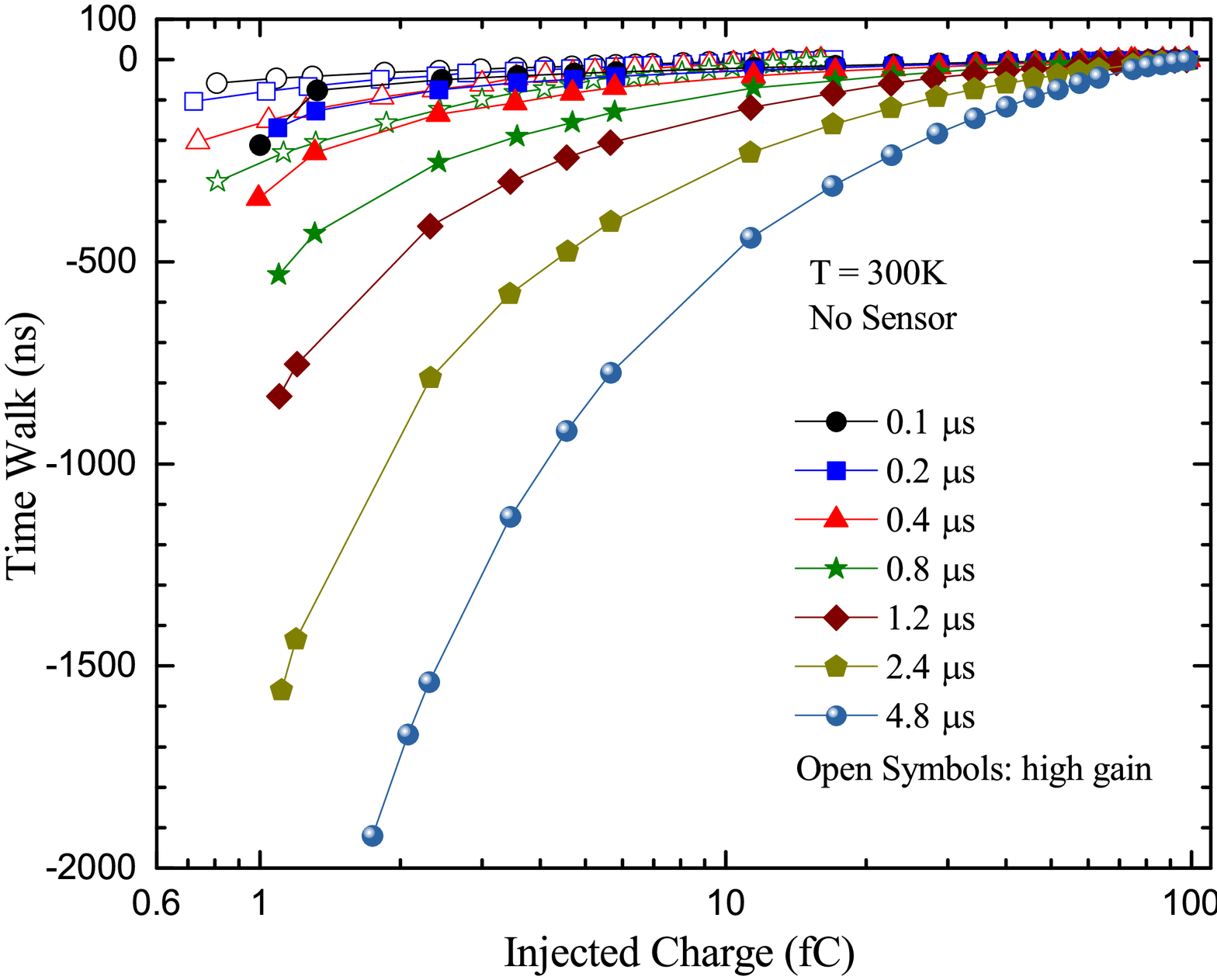}
	\caption{Measured cathode time walk at 20 mV/fC (solid symbols) and 60 mV/fC (open) symbols for injected charge up to 100 fC.}
	\label{fig:avg2CathodeTimeWalk}
\end{figure}

The cathode timing response for the threshold crossing of a fast unipolar pulse was evaluated with similar procedures to the anode except for the injection of holes into the front with the on-chip test generator. It should be noted that the high gain in the cathode was 60 mV/fC while the low gain was 20 mV/fC. At high gain and 0.25 \si{\micro\second}, charges as low as 0.8 fC were resolved at 5.4 ns with corresponding timewalk of 58 ns as shown in Figures \ref{fig:avg2CathodeTimeRes} and \ref{fig:avg2CathodeTimeWalk} respectively. At low gain and 12 \si{\micro\second} shaping time, the resolution achieved was 50 ns with a corresponding timewalk of 1.67 \si{\micro\second}. This general trend was explained in \cite{vernon:2009nss}. At low shaping time, high gain, and large charge injection, the slope of the pulse is relatively high which is ideal for good threshold crossing. Conversely, as the shaping time increased, the gain decreased and the injected charge decreased, the slope of the pulse decreased accordingly and the timewalk increased.  

The first AVG ASIC prototype (with 36 anode channels and 9 cathode channels) was used to acquire spectra with four groups of VFG CZT sensors with different geometries. Three of the sensors had a common cross-sectional area of 6 x 6 \si{\square\milli\meter} but differ in thicknesses of 20 mm, 30 mm and 40 mm while the fourth had a cross-section of 5 x 5 \si{\square\milli\meter} and was 50 mm thick. Depending on the thickness, the sensors were biased from the cathode side between 3 kV and 9 kV while the anode side was held at virtual ground by the ASIC front-end to set up the drift field inside the sensor. The measurements were taken inside an environmental chamber where the temperature was held between 17-18 \si{\degreeCelsius}. An uncollimated $^{137}$Cs was used to irradiate the sensors from $\approx$ 1 \si{\centi\meter} above the cathode side. Initially, the VFG pad sensing electrodes were connected to the ASIC anode channels for readout but it was observed that the channel's negative amplitude response to induced transients was clipped as discussed in section \ref{Anode and Pad Sensing Channels}. This issue was addressed by connecting the VFG pad electrodes to spare cathode channels for signal readout.

With this configuration, we measured between 3.2 keV to 3.5 keV for the total cathode channel noise contribution which corresponds to 0.5 \si{\percent} - 0.6 \si{\percent} FWHM at 662 keV. For each sensor, after 3D correction, we achieved 0.76 \si{\percent}, 1.06 \si{\percent}, 1.10 \si{\percent}, and 1.6 \si{\percent} FWHM at 662 keV for the 20 mm, 30 mm, 40 mm, and 50 mm thick VFG CZT bar detectors respectively. Further improvements were realized by mapping the non-uniformities of each sensor then correcting the spectra based on the uniform regions of the sensor. From this optimization, we reported improved resolutions ranging from 0.7 \si{\percent}, 0.83 \si{\percent}, 0.85 \si{\percent} and 1.5 \si{\percent} FWHM at 662 keV for the uniform regions of each sensor\cite{Bolotnikov:2016apl}. The lower resolution in the 50 mm sensor was attributed to the weak electric field near the anode.

The spectroscopic performance of the second AVG ASIC prototype (48 channels configurable as either anode or pad sensing channels and 4 dedicated cathode channels) was evaluated with a 4 x 4 array of 6 x 6 x 20 \si{\cubic\milli\meter} standard grade CZT crystals with VFG electrodes added. The sensor array was biased at 2.8 kV and irradiated at $\approx$ 3 \si{\centi\meter} from the cathode side with uncollimated $^{232}$U and $^{137}$Cs sources in an environmental chamber maintained at 23 \si{\degreeCelsius}. Using the conventional 1D (cathode/anode) correction, we reported energy resolutions of 2.8 \si{\percent} at 238 keV and 2.5 \si{\percent} at 662 keV. By applying the 3D correction it was shown that the resolution improved to 1.8 \si{\percent} at 238 keV and 0.9 \si{\percent} at 662 \cite{Ocampo:2018nim,Bolotnikov:2018nim}. Even though the leakage current and crystal defects limited the resolution, the results are very encouraging for standard grade CZT crystals that can be tiled to form commensurable large volume radiation detection systems.

\section{Conclusions and Future Work }
\label{Conclusions and Future Work }
We demonstrated an ASIC with an electronic noise contribution of less than 100 \si{\electron} at 1 \si{\micro\second} shaping time in the 3 MeV energy range of CZT. The bi-parametric measurement of amplitude and timing in each channel for above threshold incident photon events enabled the 3D correction of the spectra taken with VFG CZT bard sensors. The ASIC demonstrated that resolution less than 1 \si{\percent} FWHM at 662 keV can be obtained from spectral measurements with VFG sensors up to 40 mm thick. 

In the current design, the anode test pulse generator was only capable of injecting a maximum charge of 50 fC instead of the desired 100 fC into the anode front-end channels. This issue was attributed to a coarse scaling factor in the charge delivered by the test generator. During the evaluation of the chip, this observation was reproduced in simulation and subsequently solved. The improved layout will be implemented and verified in the next ASIC fabrication.  

Further, two other design related issues were observed. First, the anode channels and pad sensing channels shared the same threshold voltage for the respective positive pulse amplitude discriminator. During normal ASIC operation with the VFG sensors, some channels were configured as anodes while others were configured as pad sensing. In both arrangements, the positive pulse amplitude discriminators shared a common threshold DAC. This created an issue since the baselines for anode channels were $\approx$ 250 mV while that for the pad sensing channels were $\approx$ 1.2 V. After careful investigations, it was observed that setting a threshold voltage (about 3$\sigma$) above the noise floor of the 250 mV baseline resulted in normal chip operation. Under this threshold setting, the pad channels operated as designed but their shaping time had to be greater than or equal to that of the anode shaping time. Likewise, the chip performed as intended as only an anode could trigger a readout.

Second, at threshold voltages about 1 V and higher, the readout was triggered by above threshold events related to the positive pad channel amplitudes but these data sets were identifiable by their timing signatures. It is worth mentioning that a threshold this high on the anode is not practical since low energy events would be completely rejected. To address these two issues, separate DACs will be implemented for anode and pad signal discrimination. In addition, a circuit has been developed to block the flag generated by the pad signal from triggering the readout of the chip. These improvements will be implemented to ensure the full independence of each channel on the ASIC.

%% The Appendices part is started with the command \appendix;
%% appendix sections are then done as normal sections
%% \appendix

%% \section{}
%% \label{}

%% References
%%
%% Following citation commands can be used in the body text:
%% Usage of \cite is as follows:
%%   \cite{key}          ==>>  [#]
%%   \cite[chap. 2]{key} ==>>  [#, chap. 2]
%%   \citet{key}         ==>>  Author [#]

%% References with bibTeX database:

\section*{Acknowledgements}
This work was supported by the U.S. Department of Energy, Office of Defense Nuclear Nonproliferation Research $\&$ Development, DNN R$\&$D. The manuscript has been authored by Brookhaven Science Associates, LLC under Contract No. DE-AC02-98CH1-886 with the U.S. Department of Energy.

\section*{References}
\bibliographystyle{model1-num-names}
\bibliography{avg.bib}

%% Authors are advised to submit their bibtex database files. They are
%% requested to list a bibtex style file in the manuscript if they do
%% not want to use model1-num-names.bst.

%% References without bibTeX database:

% \begin{thebibliography}{00}

%% \bibitem must have the following form:
%%   \bibitem{key}...
%%

% \bibitem{}

% \end{thebibliography}

\end{document}